\documentclass[preprint,eqsecnum,aps,keywords,showpacs]{revtex4}
\usepackage{dcolumn}
\usepackage{graphicx}
\usepackage{amsmath}
\usepackage{amssymb}
\usepackage{epsfig}
\usepackage{float}
\usepackage{latexsym}
\usepackage{rotating}
\usepackage{amsmath}

\begin{document}

\title{Dyonic string-like solution in a non-Abelian gauge theory with two potentials}
\author{
Buddhi Vallabh Tripathi\footnote{Electronic address: {\em buddtrip@rediffmail.com}}${}^{}$}
\affiliation{Department of Physics, Gurukul Kangri Vishwavidyalaya, Haridwar 249 407, India}
\affiliation{Department of Physics, APB Government PG College, Agastyamuni 246 421, India}

\author{
Hemwati Nandan\footnote{Electronic address: {\em hnandan@iucaa.ernet.in}}${}^{}$}
\affiliation{Department of Physics, Gurukula Kangri Vishwavidyalaya, Haridwar 249 407, India}

\author{
K D Purohit\footnote{Electronic address: {\em kdpurohit@rediffmail.com}}${}^{}$}
\affiliation{Department of Physics, HNB Garhwal University, Srinagar Garhwal 246 174, India}

\begin{abstract}

Axially symmetric dyon solutions of a non-Abelian gauge theory model with two potentials are sought. While seeking axially symmetric (flux tube like solutions) for the model, we stumbled upon an exact solution which represents an infinite string-like dyonic configuration with cylindrical symmetry.
\\
\\
\hspace{-0.5cm}Keywords: Magnetic monopoles, Gauge field theories.
\end{abstract}

\maketitle

\section{Introduction}
Motivated by the existing (partial) duality symmetry between electric and magnetic fields in Maxwell's electrodynamics, Dirac \cite{Dirac1} introduced the idea of monopoles. The existence of the field-theoretic counterparts of the classical Dirac monopole, in certain  spontaneously broken non-Abelian (classical) gauge theories, was established by 't~Hooft \cite{tHooft1} and Polyakov \cite{Polyakov1}, independently. Julia and Zee \cite{JuliaZee1} extended their results to show the presence of dyons in a $SU(2)$ non-Abelian gauge theory with Higgs field. In fact, monopole and dyon solutions exist in any spontaneously-broken grand unified model (with the $U(1)_{EM}$ group as an embedded subgroup) \cite{tHooft1}.

To avoid the presence of singularities, which otherwise plague the conventional (single potential) theories of monopoles and dyons, theories with two electromagnetic four-vector potentials were formulated \cite{Schwinger, Zwanziger}. A two potential Abelian theory for dyons was developed by Cabibbo and Ferrari \cite{CabibboFerrari} and later Benjwal and Joshi \cite{BenjwalJoshi} extended this two potential approach for the non-Abelian case. The two potential approach has also been employed by Singleton \cite{Singleton} for a singularity free symmetric formulation of electrodynamics with magnetic charge. Recently, Singh and Tripathi \cite{SinghTripathi} have extended the Singleton's formulation for a non-Abelian two potential model to derive dyonic solutions with charges (both electric and magnetic) of topological origin.

In the present paper, we employ a set of ansatz for the gauge and scalar fields \cite{Dzhunushaliev} to seek new classical solutions for the above mentioned non-Abelian gauge theory model wih two potentials \cite{SinghTripathi} that possess axial symmetry. We have found a solution set that exists for arbitrary values of the coupling constant ($\eta$) and represents an infinitely long straight string carrying uniform (both) electric and magnetic charge densities.  
\section{The Lagrangian and field equations}
The Lagrangian density for the non-Abelian two potential model that we consider is \cite{SinghTripathi}
\begin{eqnarray}\label{Lab1}
\mathcal{L}=-\frac{1}{4}A_{\mu\nu}^a A^{\mu\nu a}+\frac{1}{4}\widetilde B_{\mu\nu}^a \widetilde B^{\mu\nu a} +\frac{1}{2}(D_\mu^1\phi_e^a)(D^{1\mu}\phi_e^a)
+\frac{1}{2}(D_\mu^2\phi_g^a)(D^{2\mu}\phi_g^a)-V(\phi_e^a,\phi_g^a),
\end{eqnarray}
where the field tensors (and duals) as well as the gauge covariant derivatives are given by
\begin{eqnarray}\label{Lab2}
A_{\mu\nu}^a&=&\partial_\mu A_\nu^a-\partial_\nu A_\mu^a-ef^{abc}A_\mu^b A_\nu^c,\\
B_{\mu\nu}^a&=&\partial_\mu B_\nu^a-\partial_\nu B_\mu^a-g f^{abc}B_\mu^b B_\nu^c,\\
D^1_\mu\phi_e^a&=&\partial_\mu\phi_e^a-e f^{abc}A_\mu^b\phi_e^c,\\
D^2_\mu\phi_g^a&=&\partial_\mu\phi_g^a-g f^{abc}B_\mu^b\phi_g^c,\\
\widetilde B_{\mu\nu}^a&=&\frac{1}{2}\varepsilon_{\mu\nu\rho\sigma}B^{\rho\sigma a},
\end{eqnarray}
with $e$ and $g$ being the coupling constants. The potential energy has the form $V(\phi_e^a,\phi_g^a)=+\eta(\phi_e^a\phi_e^a+\phi_g^a\phi_g^a-\xi^2)^2$ where $\eta$ and $\xi$ are real constants with $\eta\geq 0$. The gauge group under consideration is $SU(2)$ for which the structure constants $f^{abc}=\varepsilon^{abc}$. Due to this explicit form of the structure constants for $SU(2)$, the field tensors and the gauge covariant derivatives can be expressed as
\begin{eqnarray}
{\bf\bar A}_{\mu\nu}&=&\partial_\mu {\bf\bar A}_\nu-\partial_\nu {\bf\bar A}_\mu - e({\bf\bar A}_\mu\times {\bf\bar A}_\nu),\\ 
D^1_\mu{\bf\bar\Phi}_e&=&\partial_\mu{\bf\bar\Phi}_e-e\ {\bf\bar A}_\mu \times {\bf\bar\Phi}_e,\\
{\bf\bar B}_{\mu\nu}&=&\partial_\mu {\bf\bar B}_\nu-\partial_\nu {\bf\bar B}_\mu - g({\bf\bar B}_\mu\times {\bf\bar B}_\nu),\\ 
D^2_\mu{\bf\bar\Phi}_g&=&\partial_\mu{\bf\bar\Phi}_g-g\ {\bf\bar B}_\mu \times {\bf\bar\Phi}_g.
\end{eqnarray}
Note that all these vectors (represented in bold with overbar) and cross-products are in internal space.
The field equations corresponding to the Lagrangian density (\ref{Lab1}) are 
\begin{eqnarray}\label{Lab4}
D^1_\mu {\bf\bar A}^{\mu\nu}-e({\bf\bar \Phi}_e\times D^{1\nu}{\bf\bar\Phi}_e)&=&0,\label{Lab4.1}\\
D^2_\mu {\bf\bar B}^{\mu\nu}-g({\bf\bar \Phi}_g\times D^{2\nu}{\bf\bar\Phi}_g)&=&0,\label{Lab4.2}\\
D^1_\mu(D^{1\mu}{\bf\bar\Phi}_e)+4\eta({\bf\bar\Phi}_e^2+{\bf\bar\Phi}_g^2-\xi^2){\bf\bar\Phi}_e&=&0,\label{Lab4.3}\\
D^2_\mu(D^{2\mu}{\bf\bar\Phi}_g)+4\eta({\bf\bar\Phi}_e^2+{\bf\bar\Phi}_g^2-\xi^2){\bf\bar\Phi}_g&=&0\label{Lab4.4}.
\end{eqnarray}
We invoke the following ansatz for the four-potentials and the scalar fields \cite{Dzhunushaliev}
\begin{eqnarray}\label{ANSATZ}
{\bf\bar A}_t &=& \bigg(\frac{f_1(\rho)}{e},\ 0,\ 0\bigg), \\
{\bf\bar A}_\rho &=& 0,\\
{\bf\bar A}_\theta &=& \bigg(0,\ 0,\ \frac{\rho w_1(\rho)}{e}\bigg),\\
{\bf\bar A}_z &=& \bigg(0,\ \frac{v_1(\rho)}{e},\ 0\bigg),\\
{\bf\bar \Phi}_e &=& \bigg(\frac{\phi_1(\rho)}{e},\ 0,\ 0\bigg),
\end{eqnarray}
with a similar set of ansatz for the four-potential ${\bf\bar B_\mu}$ and scalar field ${\bf\bar \Phi}_g$ (the subscript $1\rightarrow 2$ and $e\rightarrow g$).
Here ${\bf\bar A}_t,\ {\bf\bar A}_\rho,\ {\bf\bar A}_\theta$ and ${\bf\bar A}_\phi$ are the co-variant components of the four-potential ${\bf\bar A}_\mu$ in cylindrical coordinates. Using these ansatz in eqs. (\ref{Lab4.1}) and (\ref{Lab4.3}), one obtains the following four equations:
\begin{eqnarray}\label{FVWEQN}
{f_1}^{''}+\frac{{f_1}^{'}}{\rho}&=&f_1({v_1}^2+{w_1}^2),\label{FVWEQN1}\\
{w_1}^{''}+\frac{{w_1}^{'}}{\rho}&=&w_1(-{f_1}^2+{v_1}^2-{\phi_1}^2+\frac{1}{\rho^2}),\label{FVWEQN2}\\
{v_1}^{''}+\frac{{v_1}^{'}}{\rho}&=&v_1(-{f_1}^2+{w_1}^2-{\phi_1}^2),\label{FVWEQN3}\\
{\phi_1}^{''}+\frac{{\phi_1}^{'}}{\rho}&=&\phi_1[{v_1}^2+{w_1}^2-4\eta(\frac{{\phi_1}^2}{e^2}+\frac{{\phi_2}^2}{g^2}-\xi^2)].\label{FVWEQN4}
\end{eqnarray}
Similar diffrential equations are obtained for $f_2,v_2,w_2$ and $\phi_2$, when the ansatz are plugged in eqs. (\ref{Lab4.2}) and (\ref{Lab4.4}).
\section{The dyonic string-like solution}
An exact solution set, valid for all values of the coupling constant $\eta$, for the eqs. (\ref{FVWEQN1})--(\ref{FVWEQN4}) (and their correspoding counterparts for $f_2, w_2$ \textit{etc.}) is:
\begin{eqnarray}\label{solset}
&f_1=c_1 \ln\rho, \qquad\qquad f_2=c_2 \ln \rho,&\label{solset1}\\
&v_1=w_1=0, \qquad\qquad v_2=w_2=0,&\label{solset2}\\
&\phi_1=e\xi \sin\gamma, \qquad\qquad \phi_2=g \xi\cos\gamma,&\label{solset3}
\end{eqnarray}
where $c_1, c_2$ and $\gamma$ are arbitrary constants.

In order to evaluate the electric and magnetic charges of this solution set, we have to use the 't~Hooft tensors \cite{SinghTripathi, tHooft1}
\begin{eqnarray}\label{HOFTEN1}
\mathcal{F}_{\mu\nu}={\bf\widehat\Phi}_e\cdot{\bf\bar A}_{\mu\nu} -\frac{1}{e\ {|{\bf\bar\Phi}_e|}^3}{\bf\bar\Phi}_e\cdot[ D^1_\mu {\bf\bar\Phi}_e\times D^1_\nu {\bf\bar\Phi}_e],\\
\mathcal{G}_{\mu\nu}={\bf\widehat\Phi}_g\cdot{\bf\bar B}_{\mu\nu} -\frac{1}{g\ {|{\bf\bar\Phi}_g|}^3}{\bf\bar\Phi}_g\cdot[ D^2_\mu {\bf\bar\Phi}_g\times D^2_\nu {\bf\bar\Phi}_g],
\end{eqnarray}
where ${\bf\widehat\Phi}_e={{\bf\bar\Phi}_e}/{|{\bf\bar\Phi}_e|}$ and ${\bf\widehat\Phi}_g={{\bf\bar\Phi}_g}/{|{\bf\bar\Phi}_g|}$. Now since $D_\mu^1{\bf\bar\Phi}_e=0=D_\mu^2{\bf\bar\Phi}_g$ for the solution set in eqs. (\ref{solset1})--(\ref{solset3}), the 't~Hooft tensors simply reduce as $\mathcal{F}_{\mu\nu}={\bf\widehat\Phi}_e\cdot{\bf\bar A}_{\mu\nu}$ and $\mathcal{G}_{\mu\nu}={\bf\widehat\Phi}_g\cdot{\bf\bar B}_{\mu\nu}$. The only independent non-zero components of these tensors turn out to be
\begin{eqnarray}\label{HOFTEN2}
\mathcal{F}_{t\rho}=-\frac{c_1}{e\rho},\qquad\quad \mathcal{G}_{t\rho}=-\frac{c_2}{g\rho}.
\end{eqnarray}

The components of the electric field of this solution set (eqs. (\ref{solset1})--(\ref{solset3})) are now evaluated as $E_i=\mathcal{F}^{it}-\widetilde{\mathcal{G}}^{it}$ \cite{SinghTripathi},
which leads to
\begin{eqnarray}
E_\rho&=&-\frac{c_1}{e \rho},\label{efield}\\
E_\theta &=&0,\\
E_z &=& 0.
\end{eqnarray}
Similary the magnetic field given by $B_i=\mathcal{G}^{it}+\widetilde{\mathcal{F}}^{it}$ \cite{SinghTripathi} has the components
\begin{eqnarray}
B_\rho&=&-\frac{c_2}{g \rho},\\
B_\theta &=&0,\\
B_z &=&0.\label{mfield}
\end{eqnarray}
Thus, the solutions in eqs. (\ref{solset1})--(\ref{solset3}) carry non-zero electric and magnetic charges and, hence, are dyonic. Both electric and magnetic field are cylindrically symmetric and have a ${1}/{\rho}$ dependence. Such an electric field (and magnetic field) is caused by an infinitely long uniformly charged wire (for instance see \cite{EDGRIFFITHS}).

In view of the eqs. (\ref{efield})--(\ref{mfield}), the solutions in eqs. (\ref{solset1})--(\ref{solset3}) represents an infinite string-like uniformly charged dyonic configuration.
\section{Conclusions}
Exact dyonic solutions with an infinite string-like configuration have been obtained for a non-Abelian gauge theory model with two potentials. The obtained solutions exist for all values of the coupling constant ($\eta$) in contrast to the known exact monopole and dyonic solutions, in other models, which are found only in the limit of vanishing coupling constant ($\eta\rightarrow 0$). We believe this is the only cylindrically symmetric string-like dyonic solution, obtained so far for any model of classical gauge theory. For more comprehensive understanding of other axially symmetric dyonic solutions, we intend to look for the numerical solutions of eqs. (\ref{FVWEQN1})--(\ref{FVWEQN4}) along with their relevance to the confinement problem in non-Abelian gauge theories.
\acknowledgments
HN and BVT would like to thank IUCAA, Pune for support under visiting associateship program where a part of this work was done. HN would also like to thank the Department of Science and Technology (DST), New Delhi for financial assistance through SR/FTP/PS-31/2009.

\end{document}